\documentclass[aps,prl,twocolumn]{revtex4}

\usepackage{graphicx,latexsym,color}

\def\be{\begin{equation}}
\def\ee{\end{equation}}
\def\bea{\begin{eqnarray}}
\def\eea{\end{eqnarray}}
\def\bi{\begin{itemize}}
\def\ei{\end{itemize}}
\def\K{$^{40}$K}
\def\Rb{$^{87}$Rb}
\def\exp#1{\times 10^{#1}}
\def\ket#1{\left|#1\right\rangle}

\begin{document}
\hbadness = 10000

\title{\K~\Rb~ Feshbach Resonances: Modeling the interatomic potential}

\author{C. Klempt, T. Henninger, O. Topic, J. Will, W. Ertmer, E. Tiemann and J. Arlt}
\affiliation{Institut f\"ur Quantenoptik, Leibniz Universit\"at Hannover,
Welfengarten 1, D-30167 Hannover, Germany}

\date{\today}

\begin{abstract}
We have observed 28 heteronuclear Feshbach resonances in 10 spin
combinations of the hyperfine ground states of a \K~\Rb~ mixture.
The measurements were performed by observing the loss rates from an
atomic mixture at magnetic fields between $0$ and $700$~G. This data
was used to significantly refine an interatomic potential derived
from molecular spectroscopy, yielding a highly consistent model of
the \K~\Rb~ interaction. Thus, the measured resonances can be
assigned to the corresponding molecular states. In addition, this
potential allows for an accurate calculation of the energy
differences between highly excited levels and the rovibrational
ground level. This information is of particular relevance for the
formation of deeply bound heteronuclear molecules. Finally, the
model is used to predict Feshbach resonances in mixtures of \Rb~
combined with $^{39}$K or $^{41}$K.
\end{abstract}

\maketitle One of the current challenges in modern atomic and
molecular physics is the production of quantum degenerate, dilute
ensembles with dipolar interaction. Especially for the case of
ultracold dipolar fermions, a large variety of interesting phenomena
has been predicted, including superfluid
pairing~\cite{Baranov2004,Baranov2002}, quantum
computing~\cite{DeMille2002,Micheli2006} and the fractional quantum
Hall effect~\cite{Baranov2005}. Recently, first evidence of dipolar
interactions in bosonic chromium atoms has been
reported~\cite{Stuhler2005}. However, most of the predicted effects
will be difficult to observe in atomic systems, since the dipolar
interaction is typically dominated by the contact interaction. A
fascinating alternative is the preparation of quantum degenerate
samples of molecules in their absolute ground
state~\cite{Kotochigova2003}. One obvious approach is direct cooling
of molecules (\cite{Doyle2004,Dulieu2006} and references therein).
Alternatively, weakly bound Feshbach molecules~\cite{Herbig2003, Durr2004,
Thalhammer2006, Jochim2003, Regal2003, Papp2006, Ospelkaus2006c} can
be created from degenerate atomic ensembles. Subsequently a second
process, such as Stimulated Raman Adiabatic
Passage~\cite{Shore1992}, can be used to form tightly bound
molecules~\cite{Damski2003,Kotochigova2004,Stwalley2004,Winkler2007}.
The mixture of \K~ and \Rb~ is a promising candidate for the
production of polar molecules, since the first heteronuclear
Feshbach molecules were recently obtained in this
mixture~\cite{Ospelkaus2006c} and pathways to deeply bound molecules
are under investigation~\cite{Wang2007}. The precise understanding
of the interatomic potential presented in this letter is of vital
importance for the development of such paths towards deeply bound
molecules.

In our experiments, 20 previously unknown s-wave Fesh\-bach
resonances in stable spin configurations of \K~  and \Rb~ were
detected and identified using the following procedure. An initial
prediction of the resonances was based on precise potentials of the
X$\:^{1}\Sigma^{+}$ and a$\:^{3}\Sigma^{+}$ states obtained from
molecular spectroscopy \cite{Pashov}. The asymptotic parts of these
potentials were then adjusted to reproduce the 8 previously known
resonances~\cite{Ferlaino2006,Ferlaino2006a,Ospelkaus2006a,Inouye2004}. Our
measurement of a total of 28 resonance positions as described below
allowed for a significant refinement of the potential. This
technique provides a potential which is precise both at small and
large internuclear separation. Thus, it allows for the calculation
of the energy difference between the Feshbach molecules and deeply
bound molecules in the ground states. This difference is the key
parameter for the production of deeply bound quantum degenerate
molecules with large dipole moments.

\begin{figure}[h!]

\centering

\includegraphics*[width=8.6cm]{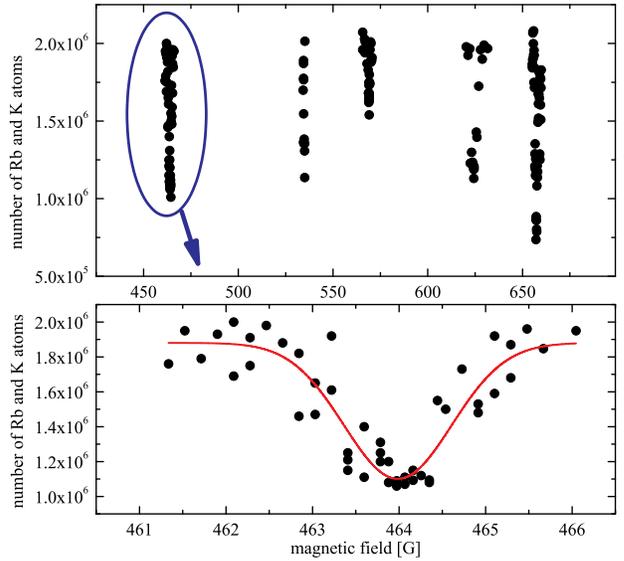}

\caption{Total atom number as a function of the magnetic field in
the \Rb~ $\ket{1,1}$, \K~ $\ket{9/2,-5/2}$ state.}

\label{fig:resonances}

\end{figure}

The following experimental procedure was used to detect
heteronuclear Feshbach resonances in a \K~\Rb~ mixture. A
two-species magneto-optical trap (MOT) was loaded with $5\exp{9}$
\Rb~ atoms and $2\exp{8}$ \K~ atoms using light-induced atom
desorption at $395$~nm. This process provides an efficient,
switchable source of atoms~\cite{Klempt2006}. The desorption light
was switched off before the end of the MOT phase in order to benefit
from a better vacuum. Both species were cooled in an optical
molasses and optically pumped to the fully stretched low-field
seeking Zeeman states, $\ket{f,m_f} = \ket{2,2}$ for \Rb~ and
$\ket{9/2,9/2}$ for \K. The atoms were subsequently loaded into a
magnetic quadrupole trap and mechanically transported to an
ultra-high vacuum cell. There, these atoms were transferred into a
harmonic trap in QUIC configuration~\cite{Esslinger1998} with \Rb~
trap frequencies of $2 \pi \times 230$~Hz ($2 \pi \times 23$~Hz) in
radial (axial) direction. The \Rb~ atoms were cooled by forced radio
frequency (RF) evaporation and in turn cooled the \K~ atoms
sympathetically. This evaporation was terminated shortly before
reaching quantum degeneracy and an equal mixture of $2 \exp{6}$ \Rb~
and \K~ atoms at a temperature of $1.1~\mu$K with no discernible
fraction in other Zeeman states was obtained. This ensemble was
transferred into an optical dipole trap formed by a monolithic
Nd:YAG laser at a wavelength of $1064$~nm with a trap depth of
$170~\mu$K and radial (axial) trapping frequencies of $2 \pi \times
2$~kHz ($2 \pi \times 25$~Hz) for \Rb. In the dipole trap the
mixture had a temperature of $4~\mu$K and a maximum \Rb~ density of
$2 \exp{14}$~atoms/cm$^3$.

\begin{table}[!ht]
\begin{center}
\caption{Magnetic-field positions and widths of all investigated
s-wave Feshbach resonances in the mixture of \K~ and \Rb. Quantum
numbers are given for the corresponding molecular levels without
magnetic field. The whole multiplet is given for the total angular
momentum $F$, since its splitting is smaller than the Zeeman energy
(see Fig. \ref{fig:E_vs_B}) and $F$ is no longer a good quantum
number. All molecular states are mixed levels of the last bound
vibrational states $v_{\rm a}=30$ and $v_{\rm X}=98$. Note that the
experimental width $\sigma_{\rm exp}$ does not correspond directly
to the theoretical width $\Delta_{\rm th}$ (see definition in
text).} \label{tab:list} \vskip 3pt

\begin{tabular}{l | c  c | c  c | c | c | c }
\hline \hline

 $m_{\rm Rb},$& $B_{\rm th}$  & $-\Delta_{\rm th}$  & $B_{\rm exp}$&$\sigma_{\rm exp}$ & $F$ & $f_{\rm Rb}/$& \\
$m_{\rm K}$& (G) & (G)  & (G) & (G) &  &$f_{\rm K}$ &  \\

\hline
\hline

$1$,$\;\frac{7}{2}$& 
$298.70$&$0.61$&$298.67$&$<0.12$  &$\frac{9}{2},\frac{11}{2}$&$1/\frac{9}{2}$&\scriptsize \cite{Ferlaino2006a}\\[1pt]
\hline
$1$,$\;\frac{5}{2}$& 
$177.57$&$<0.005 $&$  -  $&$  - $ &$\frac{7}{2},\frac{9}{2},\frac{11}{2}$&$1/\frac{9}{2}$&\\
&$359.86$&$ 0.88 $&$ 359.70 $&$ <0.1 $ &$\frac{7}{2},\frac{9}{2},\frac{11}{2}$&$1/\frac{9}{2}$&\\[1pt]
\hline
$1$,$\;\frac{3}{2}$& 
$184.56$&$<0.005$&$  -  $&$  - $ &$\frac{7}{2},\frac{9}{2},\frac{11}{2}$&$1/\frac{9}{2}$&\\
&$399.12$&$ 0.85 $&$ 399.16 $&$ 0.61 $ &$\frac{7}{2},\frac{9}{2},\frac{11}{2}$&$1/\frac{9}{2}$&\\[1pt]
\hline
$1$,$\;\frac{1}{2}$&$ 
190.25 $&$<0.005$&$  -  $&$  -  $ &$\frac{7}{2},\frac{9}{2},\frac{11}{2}$&$1/\frac{9}{2}$&\\
&$ 424.37 $&$ 0.76 $&$ 424.39 $&$ 0.30 $ &$\frac{7}{2},\frac{9}{2},\frac{11}{2}$&$1/\frac{9}{2}$&\\
&$ 660.15 $&$ 0.11 $&$ 660.23 $&$ 0.42 $ &$\frac{3}{2}\ldots\frac{11}{2}$&$2/\frac{7}{2}$&\\[1pt]
\hline
$1$,$\;-\frac{1}{2}$&$ 
194.90 $&$<0.005$&$  -  $&$  -  $ &$\frac{7}{2},\frac{9}{2},\frac{11}{2}$&$1/\frac{9}{2}$&\\
&$ 441.87 $&$ 0.66 $&$ 442.06 $&$ 0.55 $ &$\frac{7}{2},\frac{9}{2},\frac{11}{2}$&$1/\frac{9}{2}$&\\
&$ 612.69 $&$ 0.08 $&$ 612.48 $&$ <0.24 $ &$\frac{3}{2}\ldots\frac{11}{2}$&$1/\frac{7}{2}$&\\
&$ 688.55 $&$ 0.01 $&$ 688.56 $&$ 0.34 $ &$\frac{5}{2},\frac{7}{2},\frac{9}{2}$&$2/\frac{7}{2}$&\\[1pt]
\hline
$1$,$\;-\frac{3}{2}$&$ 
198.74 $&$<0.005$&$  -  $&$  -  $ &$\frac{7}{2},\frac{9}{2},\frac{11}{2}$&$1/\frac{9}{2}$&\\
&$ 454.66 $&$ 0.55 $&$ 454.88 $&$ 0.82 $ &$\frac{7}{2},\frac{9}{2},\frac{11}{2}$&$1/\frac{9}{2}$&\\
&$ 571.19 $&$ 0.05 $&$ 571.17 $&$ <0.24 $ &$\frac{3}{2}\ldots\frac{11}{2}$&$2/\frac{7}{2}$&\\
&$ 623.23 $&$ 0.02 $&$ 623.29 $&$ 0.45 $ &$\frac{5}{2},\frac{7}{2},\frac{9}{2}$&$1/\frac{7}{2}$&\\
&$ 669.87 $&$ 0.81 $&$ 669.84 $&$ <0.14 $ &$\frac{3}{2}\ldots\frac{11}{2}$&$2/\frac{7}{2}$&\\[1pt]
\hline
$1$,$\;-\frac{5}{2}$&$ 
201.96 $&$<0.005 $&$  -  $&$  -  $&$\frac{7}{2},\frac{9}{2},\frac{11}{2}$&$1/\frac{9}{2}$&\\
&$ 464.02 $&$ 0.42 $&$ 463.89 $&$ 0.62 $ &$\frac{7}{2},\frac{9}{2},\frac{11}{2}$&$1/\frac{9}{2}$&\\
&$ 534.66 $&$ 0.02 $&$ 534.68 $&$ <0.20 $ &$\frac{3}{2}\ldots\frac{11}{2}$&$2/\frac{7}{2}$&\\
&$ 568.14 $&$ 0.03 $&$ 568.28 $&$ 0.55 $ &$\frac{5}{2},\frac{7}{2},\frac{9}{2}$&$1/\frac{7}{2}$&\\
&$ 624.89 $&$ 0.75 $&$ 624.29 $&$ 1.56 $ &$\frac{3}{2}\ldots\frac{11}{2}$&$1/\frac{7}{2}$&\\
&$ 657.09 $&$ 2.66 $&$ 657.24 $&$ 0.97 $ &$\frac{5}{2},\frac{7}{2},\frac{9}{2}$&$2/\frac{7}{2}$&\\[1pt]
\hline
$1$,$\;-\frac{7}{2}$&$ 
204.65 $&$ <0.002 $&$  -  $&$  -  $ &$\frac{7}{2},\frac{9}{2},\frac{11}{2}$&$1/\frac{9}{2}$&\\
&$ 469.48 $&$ 0.28 $&$ 469.54 $&$ <0.14 $&$\frac{7}{2},\frac{9}{2},\frac{11}{2}$&$1/\frac{9}{2}$&\scriptsize \cite{Ferlaino2006a} \\
&$ 523.04 $&$ 0.06 $&$ 523.04 $&$ <0.2 $ &$\frac{5}{2},\frac{7}{2},\frac{9}{2}$&$1/\frac{7}{2}$&\\
&$ 584.58 $&$ 0.70 $&$ 584.42 $&$ 0.35 $&$\frac{5}{2}\ldots\frac{11}{2}$&$1/\frac{7}{2}$&\scriptsize \cite{Ferlaino2006a} \\
&$ 598.32 $&$ 2.55 $&$ 598.24 $&$ 0.32 $&$\frac{5}{2},\frac{7}{2},\frac{9}{2}$&$2/\frac{7}{2}$&\scriptsize \cite{Ferlaino2006a} \\
&$ 697.56 $&$ 0.16 $&$ 697.80 $&$ 0.49 $&$\frac{5}{2},\frac{7}{2},\frac{9}{2}$&$1/\frac{7}{2}$&\scriptsize \cite{Ferlaino2006a} \\[1pt]
\hline
$1$,$\;-\frac{9}{2}$&$ 
207.02 $&$ <0.001 $&$  -  $&$  -  $ &$\frac{7}{2},\frac{9}{2},\frac{11}{2}$&$1/\frac{9}{2}$&\\
&$ 462.45 $&$ 0.06 $&$ 462.41 $&$ <0.20 $ &$\frac{7}{2},\frac{9}{2},\frac{11}{2}$&$1/\frac{9}{2}$&\\
&$ 495.71 $&$ 0.14 $&$ 495.57 $&$ <0.20 $&$\frac{7}{2},\frac{9}{2}$&$1/\frac{7}{2}$&\scriptsize \cite{Ferlaino2006a,Inouye2004,Ospelkaus2006a} \\
&$ 546.89 $&$ 3.07 $&$ 546.71 $&$ 0.68 $&$\frac{7}{2},\frac{9}{2}$&$1/\frac{7}{2}$&\scriptsize \cite{Ferlaino2006a,Inouye2004,Ospelkaus2006a} \\
&$ 659.68 $&$ 0.80 $&$ 659.52 $&$ 0.39 $&$\frac{7}{2},\frac{9}{2},\frac{11}{2}$&$2/\frac{9}{2}$&\scriptsize \cite{Ferlaino2006a} \\[1pt]
\hline
$0$,$\;-\frac{9}{2}$& 
$430.71 $&$ 0.11 $&$ 430.93 $&$ 0.63 $ &$\frac{9}{2},\frac{11}{2}$&$1/\frac{9}{2}$&\\
&$ 546.08 $&$ 3.16 $&$ 546.13 $&$ 0.43 $ &$\frac{9}{2}$&$1/\frac{7}{2}$&\\[1pt]

\hline\hline

\end{tabular}
\end{center}
\vskip -10mm
\end{table}

The use of an optical dipole trap allows for the free choice of the
magnetic field and thus enables an investigation of Feshbach
resonances. To produce strong magnetic fields up to $700$~G, the
quadrupole coils of the magnetic trap are operated in
quasi-Helmholtz configuration. This field was raised to $10$~G to
transfer the \Rb~ atoms from the $\ket{2,2}$ state to the
$\ket{1,1}$ state by a microwave rapid adiabatic passage. Starting
at $6.8561$~GHz, the microwave frequency was reduced by $800$~kHz
within a $50$~ms interval. The transfer efficiency was better than
$90\%$ and the remaining atoms in the $\ket{2,2}$ state were removed
with a resonant light pulse. Afterwards, the magnetic field was
raised to $19$~G to allow for a transfer of the \K~ atoms to a
specific spin state. To access the nine
$\ket{f=9/2,m_f=-9/2\ldots7/2}$ states, radio frequency sweeps with
a speed of $10$~kHz/ms starting at $6.56$~MHz were used. The desired
spin state was selected by adjusting the end of the RF sweep.

If needed, the magnetic field was raised to $29$~G to transfer the
\Rb~ atoms to the $\ket{1,0}$ or $\ket{1,-1}$ state by using RF
sweeps. The final mixture is stable against spin changing collisions
as long as one of the components is in the lowest energy spin state.
In this case, the spin exchange is either forbidden by energy or by
angular momentum conservation, since the \Rb~ Zeeman structure is
inverted. This allows for the investigation of Feshbach resonances
in a total of 12 different spin configurations, nine of which have
not been investigated so far
\cite{Ferlaino2006,Ferlaino2006a,Inouye2004,Ospelkaus2006a}.

The Feshbach resonances in all of these states were located by
detecting inelastic losses. After preparing the spin mixture of
interest, the magnetic field was quickly increased to a certain
value. The atoms were held at this magnetic field for a time between
$0.5$~s and $2$~s. The hold time was chosen such that the maximum
loss of atoms was approximately $50\%$. On the one hand, this
ensures a good signal-to-noise ratio, on the other hand, the
population of other Zeeman states due to inelastic collisions
remains negligible. After this hold time, the magnetic field was
lowered to a small bias field perpendicular to the direction of
detection. Both species were detected independently by absorption
imaging after ballistic expansion. The spin states were separated in
a Stern-Gerlach experiment by a strong magnetic gradient applied for
$4$~ms during the initial stage of the expansion. This allowed us to
verify that no unwanted spin components were populated. The total
number of atoms was measured as a function of the applied magnetic
field. Figure~\ref{fig:resonances} shows the result of these
measurements for the spin configuration \Rb~$\ket{1,1}$ and
\K~$\ket{9/2,-5/2}$. For each resonance, a Gaussian fit was used to
extract the center position and the $1/e^2$-width of the resonance.
Due to slight asymmetries of the resonances and shot-to-shot atom
number fluctuations of $\sim 10\%$, the determination of the
resonance position has a mean relative uncertainty of $3 \exp{-4}$.

The magnetic field was calibrated at four magnetic field strengths
between $200$ and $700$~G using the \Rb~ $\ket{1,1} \rightarrow
\ket{1,0}$ RF transition. A residual inhomogeneity of the magnetic
field over the trap region and small current noise lead to a
relative field width of $3 \exp{-4}$, restricting the uncertainty of
the field calibration to $1 \exp{-4}$.

Based on the interatomic potential we were able to predict all
s-wave Feshbach resonances in the available spin states. All
resonances were found less than $0.8$~G from the predicted values.
The measured resonances are listed in Table \ref{tab:list}.

\begin{figure}[ht!]

\centering

\includegraphics*[width=8.6cm]{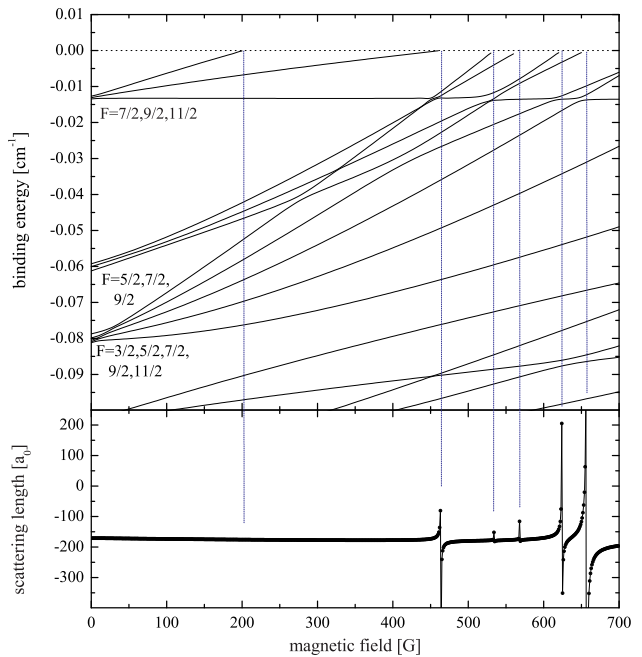}

\caption{Simulation of the molecular binding energies as a function
of the magnetic field for the \Rb~$\ket{1,1}$, \K~$\ket{9/2,-5/2}$
state (top). Around the magnetic field positions where the binding
energy approaches zero, the heteronuclear scattering length diverges
(bottom). The resonances are not fully resolved due to the finite
computational step size chosen for the graph. The total angular
momentum specified at zero magnetic field is denoted by $F$.}

\label{fig:E_vs_B}
\end{figure}

The measured resonance positions were then used to refine the
molecular potentials further using the method described in
Ref.~\cite{Pashov}. By slightly altering the dispersive coefficients
and the slope of the potential at the inner turning point, an
optimal match of the calculated and measured resonance positions was
obtained. In this procedure, the experimental observations were
weighted with their individual uncertainties. Figure
\ref{fig:E_vs_B} shows the binding energies and the scattering
length at the resonances for the \Rb~$\ket{1, 1}$, \K~$\ket{9/2,
-5/2}$ state. The calculated resonance positions $B_{\rm th}$ and
widths $\Delta_{\rm th}$ are listed in Table \ref{tab:list}. The
theoretical widths correspond to the distance between the center of
the resonance and the zero crossing of the scattering length. This
width corresponds to the usual approximation for the scattering
length in the vicinity of a Feshbach resonance $a(B) = a_{\rm BG} (1
- \frac{\Delta_{\rm th}}{B-B_{\rm th}})$ with the background
scattering length $a_{\rm BG}$.

With this procedure, it was possible to improve the consistency of
the model significantly. The initial fit to the ten known resonance
positions from Ref.~\cite{Ferlaino2006,Ferlaino2006a} gave a mean deviation of
$275$~mG, $2.75$ times larger than the experimental uncertainty of
$100$~mG. We were able to reduce the mean deviation for our
measurement of all 28 resonances to $123$~mG, yielding a standard
deviation of $1.1$ times the experimental uncertainty. Thus, our
model reproduces all resonance positions within the experimental
uncertainty. However, our model does not include the collisional
loss processes in the vicinity of a resonance which can slightly
shift the resonance positions.

The precise knowledge of the interatomic potential is a key
ingredient for many experiments with quantum degenerate
heteronuclear mixtures. In particular, our model allows for an
accurate calculation of binding energies of the molecular
levels, which is an essential ingredient on the way to deeply bound
quantum degenerate molecules. The weakly bound levels which are
responsible for the Feshbach resonances are listed in Table
\ref{tab:binding}. For the lowest singlet and triplet molecular
states, we obtain a binding energy of $4180.269$~cm$^{-1}$
(N=0, hyperfine splitting less than $1$~MHz) and $240.024$~cm$^{-1}$
(N=0, F=7/2), respectively.

\begin{table}[!ht]
\begin{center}
\caption{Binding energies of the involved molecular states in
cm$^{-1}$ with respect to the atomic asymptote $f_{\rm Rb}=1$,
$f_{\rm K}=9/2$. The expectation value of the electronic spin is
given to quantify the triplet-singlet mixing.} \label{tab:binding}
\vskip 3pt

\begin{tabular}{c | c | c | l || c | c | c | l}
\hline \hline
 $F$ & $f_{\rm Rb} / f_{\rm K}$ & spin & binding&$F$ & $f_{\rm Rb} / f_{\rm K}$ & spin & binding\\
&&& energy&&&& energy \\
\hline
$\frac{3}{2}$&$2/\frac{7}{2}$&$1$&$-0.08120$ &       $\frac{9}{2}$ &$1/\frac{7}{2}$&$0.831$&$-0.06125$ \\
$\frac{5}{2}$&$2/\frac{7}{2}$&$0.918$&$-0.08096$ &  $\frac{7}{2}$ &$1/\frac{7}{2}$&$0.718$&$-0.06009$ \\
$\frac{7}{2}$&$2/\frac{7}{2}$&$0.81$&$-0.08056$ &   $\frac{5}{2}$ &$1/\frac{7}{2}$&$0.638$&$-0.05926$ \\
$\frac{9}{2}$&$2/\frac{7}{2}$&$0.678$&$-0.07989$ &  $\frac{7}{2}$ &$1/\frac{9}{2}$&$0.907$&$-0.01345$ \\
$\frac{11}{2}$&$2/\frac{7}{2}$&$0.519$&$-0.07878$ & $\frac{9}{2}$ &$1/\frac{9}{2}$&$0.784$&$-0.01310$ \\
&&&&                                               $\frac{11}{2}$&$1/\frac{9}{2}$&$0.628$&$-0.01269$ \\[1pt]
\hline \hline
\end{tabular}
\end{center}
\end{table}

In combination with a good understanding of the electronically
excited molecular potentials, this will allow for the selection of
excited states with sufficient Franck-Condon overlap for the
transfer of Feshbach molecules to deeply bound molecules. Thus, our
results provide a major step on the path towards the production of
quantum degenerate ground-state molecules.

\begin{figure}[ht!]

\centering

\includegraphics*[width=8.6cm]{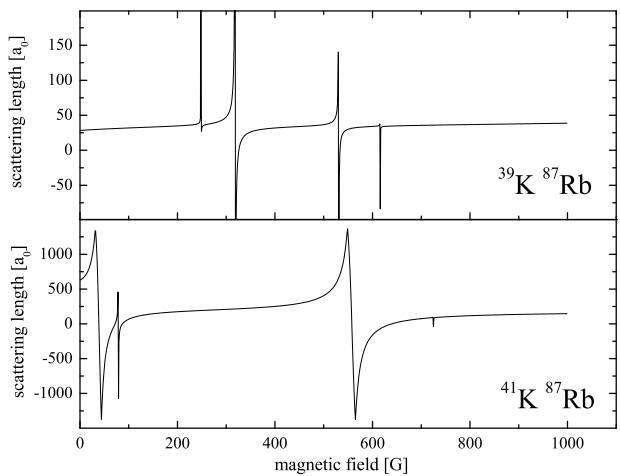}

\caption{Interspecies scattering lengths of $^{39}$K\Rb~ (top) and
$^{41}$K\Rb~ (bottom) as a function of the magnetic field strength.}

\label{fig:K41_39}

\end{figure}

Finally, our model allows us to predict Feshbach resonances in
mixtures of $^{39}$K and \Rb~ or $^{41}$K and \Rb. In Fig.
\ref{fig:K41_39}, the Feshbach resonances are shown for the stable
spin configuration with both species in the $\ket{1,1}$ state. Both
mixtures show broad resonances, which can be used for precise
control of the interactions~\cite{Roati2007}. In particular, the
wide resonance at $\sim 40~G$ will allow very precise control of the
scattering properties over the full range from $-1500~a_0$ to
$1500~a_0$. Both bosonic Potassium-Rubidium mixtures are currently
under investigation and will enable an experimental investigation of
the predicted resonances. By comparing the resonance positions, it
will be possible to test the validity of the Born-Oppenheimer
potentials used in this work. The non-zero mass ratio between the
electrons and the nuclei leads to corrections to the
Born-Oppenheimer potential~\cite{Watson1980} which result in a phase
shift of the scattering states. This shifts the Fesh\-bach
resonances slightly with respect to the predictions and an
experimental observation will allow for the validation of the
mass-scaling techniques with unprecedented precision.

In conclusion, we have observed 20 previously unknown s-wave
Feshbach resonances in a mixture of \K~ and \Rb. Using these
observations we have significantly improved the model of the
interspecies interaction, which reproduces the experimental data
both at small and large internuclear separation. In addition, this
potential allows for an accurate calculation of the energy
differences between highly excited states and the molecular ground
state. This information is of particular relevance for fascinating
prospect of forming deeply bound heteronuclear molecules. Finally,
the model is used to predict Feshbach resonances in mixtures of \Rb~
combined with $^{39}$K or $^{41}$K. These resonances promise precise
control of the bosonic mixtures and also provide a novel system to
test the validity of the Born-Oppenheimer approximation.

\bibliographystyle{prsty}
\bibliography{fesh}
\end{document}